\documentclass[ reprint, amsmath,amssymb, aps, prl, showpacs, bibnotes,superscriptaddress]{revtex4-1}

\usepackage{graphicx}
\usepackage{dcolumn}
\usepackage{bm}
\begin{document}
\title{Measurement of Atmospheric Neutrino Oscillations with IceCube}
\affiliation{III. Physikalisches Institut, RWTH Aachen University, D-52056 Aachen, Germany}
\affiliation{School of Chemistry \& Physics, University of Adelaide, Adelaide SA, 5005 Australia}
\affiliation{Dept.~of Physics and Astronomy, University of Alaska Anchorage, 3211 Providence Dr., Anchorage, AK 99508, USA}
\affiliation{CTSPS, Clark-Atlanta University, Atlanta, GA 30314, USA}
\affiliation{School of Physics and Center for Relativistic Astrophysics, Georgia Institute of Technology, Atlanta, GA 30332, USA}
\affiliation{Dept.~of Physics, Southern University, Baton Rouge, LA 70813, USA}
\affiliation{Dept.~of Physics, University of California, Berkeley, CA 94720, USA}
\affiliation{Lawrence Berkeley National Laboratory, Berkeley, CA 94720, USA}
\affiliation{Institut f\"ur Physik, Humboldt-Universit\"at zu Berlin, D-12489 Berlin, Germany}
\affiliation{Fakult\"at f\"ur Physik \& Astronomie, Ruhr-Universit\"at Bochum, D-44780 Bochum, Germany}
\affiliation{Physikalisches Institut, Universit\"at Bonn, Nussallee 12, D-53115 Bonn, Germany}
\affiliation{Universit\'e Libre de Bruxelles, Science Faculty CP230, B-1050 Brussels, Belgium}
\affiliation{Vrije Universiteit Brussel, Dienst ELEM, B-1050 Brussels, Belgium}
\affiliation{Dept.~of Physics, Chiba University, Chiba 263-8522, Japan}
\affiliation{Dept.~of Physics and Astronomy, University of Canterbury, Private Bag 4800, Christchurch, New Zealand}
\affiliation{Dept.~of Physics, University of Maryland, College Park, MD 20742, USA}
\affiliation{Dept.~of Physics and Center for Cosmology and Astro-Particle Physics, Ohio State University, Columbus, OH 43210, USA}
\affiliation{Dept.~of Astronomy, Ohio State University, Columbus, OH 43210, USA}
\affiliation{Dept.~of Physics, TU Dortmund University, D-44221 Dortmund, Germany}
\affiliation{Dept.~of Physics, University of Alberta, Edmonton, Alberta, Canada T6G 2E1}
\affiliation{D\'epartement de physique nucl\'eaire et corpusculaire, Universit\'e de Gen\`eve, CH-1211 Gen\`eve, Switzerland}
\affiliation{Dept.~of Physics and Astronomy, University of Gent, B-9000 Gent, Belgium}
\affiliation{Dept.~of Physics and Astronomy, University of California, Irvine, CA 92697, USA}
\affiliation{Laboratory for High Energy Physics, \'Ecole Polytechnique F\'ed\'erale, CH-1015 Lausanne, Switzerland}
\affiliation{Dept.~of Physics and Astronomy, University of Kansas, Lawrence, KS 66045, USA}
\affiliation{Dept.~of Astronomy, University of Wisconsin, Madison, WI 53706, USA}
\affiliation{Dept.~of Physics and Wisconsin IceCube Particle Astrophysics Center, University of Wisconsin, Madison, WI 53706, USA}
\affiliation{Institute of Physics, University of Mainz, Staudinger Weg 7, D-55099 Mainz, Germany}
\affiliation{Universit\'e de Mons, 7000 Mons, Belgium}
\affiliation{T.U. Munich, D-85748 Garching, Germany}
\affiliation{Bartol Research Institute and Department of Physics and Astronomy, University of Delaware, Newark, DE 19716, USA}
\affiliation{Dept.~of Physics, University of Oxford, 1 Keble Road, Oxford OX1 3NP, UK}
\affiliation{Dept.~of Physics, University of Wisconsin, River Falls, WI 54022, USA}
\affiliation{Oskar Klein Centre and Dept.~of Physics, Stockholm University, SE-10691 Stockholm, Sweden}
\affiliation{Department of Physics and Astronomy, Stony Brook University, Stony Brook, NY 11794-3800, USA}
\affiliation{Dept.~of Physics and Astronomy, University of Alabama, Tuscaloosa, AL 35487, USA}
\affiliation{Dept.~of Astronomy and Astrophysics, Pennsylvania State University, University Park, PA 16802, USA}
\affiliation{Dept.~of Physics, Pennsylvania State University, University Park, PA 16802, USA}
\affiliation{Dept.~of Physics and Astronomy, Uppsala University, Box 516, S-75120 Uppsala, Sweden}
\affiliation{Dept.~of Physics, University of Wuppertal, D-42119 Wuppertal, Germany}
\affiliation{DESY, D-15735 Zeuthen, Germany}
\affiliation{Max-Planck-Institut f\"ur Physik, D-80805 M\"unchen, Germany}

\author{M.~G.~Aartsen}
\affiliation{School of Chemistry \& Physics, University of Adelaide, Adelaide SA, 5005 Australia}
\author{R.~Abbasi}
\affiliation{Dept.~of Physics and Wisconsin IceCube Particle Astrophysics Center, University of Wisconsin, Madison, WI 53706, USA}
\author{Y.~Abdou}
\affiliation{Dept.~of Physics and Astronomy, University of Gent, B-9000 Gent, Belgium}
\author{M.~Ackermann}
\affiliation{DESY, D-15735 Zeuthen, Germany}
\author{J.~Adams}
\affiliation{Dept.~of Physics and Astronomy, University of Canterbury, Private Bag 4800, Christchurch, New Zealand}
\author{J.~A.~Aguilar}
\affiliation{D\'epartement de physique nucl\'eaire et corpusculaire, Universit\'e de Gen\`eve, CH-1211 Gen\`eve, Switzerland}
\author{M.~Ahlers}
\affiliation{Dept.~of Physics and Wisconsin IceCube Particle Astrophysics Center, University of Wisconsin, Madison, WI 53706, USA}
\author{D.~Altmann}
\affiliation{Institut f\"ur Physik, Humboldt-Universit\"at zu Berlin, D-12489 Berlin, Germany}
\author{J.~Auffenberg}
\affiliation{Dept.~of Physics and Wisconsin IceCube Particle Astrophysics Center, University of Wisconsin, Madison, WI 53706, USA}
\author{X.~Bai}
\thanks{Physics Department, South Dakota School of Mines and Technology, Rapid City, SD 57701, USA}
\affiliation{Bartol Research Institute and Department of Physics and Astronomy, University of Delaware, Newark, DE 19716, USA}
% ...

\collaboration{IceCube Collaboration}
\noaffiliation

\date{\today}
\begin{abstract}
We present the first statistically significant detection of neutrino oscillations in the high-energy regime ($>$ 20 GeV) 
from an analysis of IceCube Neutrino Observatory data collected in 2010-2011.
This measurement is made possible by the low energy threshold of the DeepCore detector ($\sim 20$ GeV) 
and benefits from the use of the IceCube detector as a veto against cosmic ray-induced muon background.
The oscillation signal was detected within a low-energy muon neutrino sample (20 -- 100 GeV) extracted from data collected by DeepCore. 
A high-energy muon neutrino sample (100 GeV -- 10 TeV) was extracted from IceCube data to constrain systematic uncertainties. 
Disappearance of low-energy upward-going muon neutrinos was observed, and
the non-oscillation hypothesis is rejected with more than $5\sigma$ significance. In a two-neutrino flavor 
formalism, our data are best described by the atmospheric neutrino oscillation parameters 
$\Delta m^2_{23}= (2.3^{+0.6}_{-0.5})\cdot 10^{-3}$ eV$^2$ and $\sin^2(2 \theta_{23})>0.93$, and maximum mixing is favored. 

\end{abstract}

\maketitle



Neutrino flavor oscillations are now an established fact.
A number of experiments have observed this physical phenomenon
over a wide range of energies, spanning from a fraction of MeV to several GeV.
Measurements above 10 GeV have been relatively limited because of constraints of detector volume, neutrino beam energy and/or insufficient 
distance of the detector to the beam source.
With the construction of high energy neutrino telescopes with very large volumes and abundant atmospheric neutrinos, 
studies of neutrino properties above 10 GeV have become possible. 
Recently, the ANTARES collaboration reported a first indication ($2.3 \sigma$) of 
atmospheric neutrino oscillations in the 20 GeV to 100 GeV energy band \cite{AdrianMartinez:2012ph}. 
In this paper we report the first statistically significant observation of atmospheric $\nu_{\mu}$ disappearance in this energy band.

Flavor oscillations of atmospheric neutrinos traversing the Earth are a complex process, since all three active
flavors may transform into one another. Furthermore, matter 
effects modify the effective oscillation parameters in the Earth from the vacuum values. However, 
for the energy range of this analysis, a two-flavor vacuum-like description ($\nu_\mu \rightarrow \nu_\tau$) is an adequate
approximation.
In this scenario, the muon neutrino survival probability is
\begin{equation}
P(\nu_\mu\rightarrow\nu_\mu)= 1 -\sin^2 (2\theta_{23}) \sin^2(1.27 \Delta m^2_{23}L/E) 
\end{equation}
where $\Delta m^2_{23}$ is the atmospheric mass-squared difference in eV$^2$, $\theta_{23}$ is the atmospheric mixing angle, 
$L$ is the propagation distance in km, and $E$ is the neutrino energy in GeV. 
Full numerical three-flavor calculations in matter found differences from this formula of less than a few percent.
Given the resolution of the present analysis, this approximation is sufficiently accurate. 


This analysis uses data collected from May 2010 to May 2011 by the IceCube neutrino telescope, including its low-energy sub-detector 
DeepCore \cite{Collaboration:2011ym}.
IceCube is a cubic-kilometer neutrino detector installed in the ice at
the geographic South Pole \cite{Achterberg:2006md}. Neutrino detection relies on the optical detection of Cherenkov radiation emitted by
secondary particles produced in neutrino interactions in the surrounding ice or the nearby bedrock. 
This analysis detects muons produced in charged current interactions of $\nu_\mu$ which can travel large distances in the ice. Their long tracks can be
reconstructed and provide information about the direction of the initial neutrino.
IceCube's optical sensors, Digital Optical Modules (DOMs), consist of $25.4$ cm photomultipliers tubes in a glass 
pressure housing with in-situ pulse digitization \cite{Abbasi:2008aa,Abbasi:2010vc}. 
The sensors are arranged on 86 vertical strings, each holding 60 DOMs. 
The primary (high-energy) detector has a spacing of 17 m between sensors and an average horizontal distance of 125 m 
between neighboring strings. The low-energy infill array DeepCore 
consists of eight dedicated strings with a typical spacing
of $70$ m deployed near the center of the IceCube array.
On the dedicated DeepCore strings, the sensors are concentrated in the clearest deep ice, with a denser $7$ m vertical 
spacing. This analysis uses data taken while 79 detector strings were operational (IceCube-79), including six of the 
dedicated DeepCore strings. A total of $318.9$ days of high-quality data were collected
in this configuration, excluding periods of calibration runs, partial detector configurations
and detector downtime. 

\begin{figure}[htb!]
\begin{center}
\includegraphics[width=0.4\textwidth]{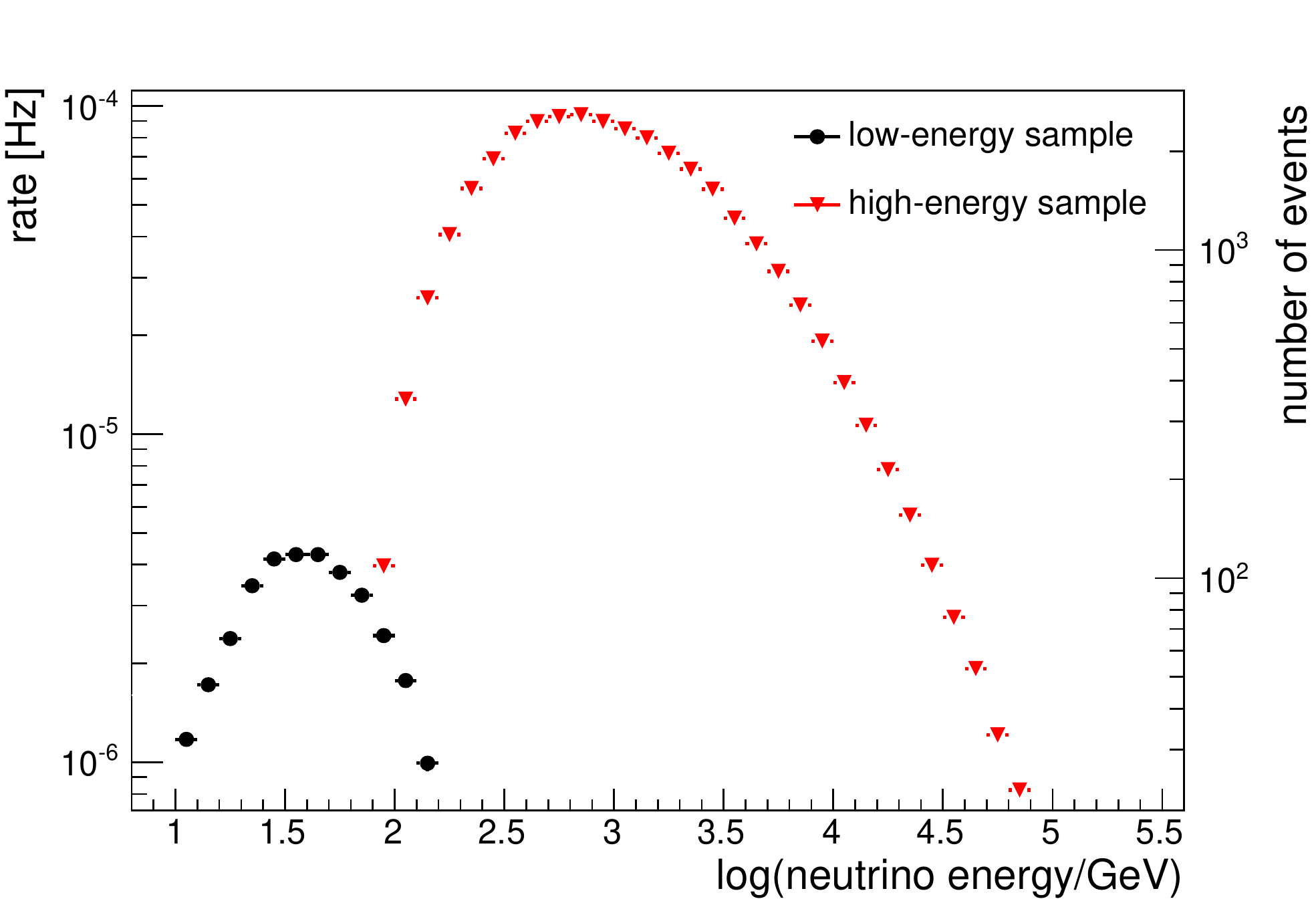}
\end{center}
\caption{\label{fig:energy}Expected distribution of the neutrino energy of atmospheric neutrinos in the low-energy (DeepCore) and in the high-energy (IceCube) 
samples according to simulations.}
\end{figure}

The aim of this analysis was to experimentally measure an expected modification of the atmospheric neutrino zenith angle 
distribution due to oscillation-induced muon neutrino disappearance. From Eq.~(1) we expected the effect to be strongest for vertical 
events with neutrino energies around 25 GeV. Two samples of upward-going muon neutrino events were extracted from data. The first sample was obtained  
from relatively high-energy events using data from the entire IceCube detector. The second sample, selected from events 
starting in the DeepCore volume, 
was very pure in lower energy neutrinos after using the surrounding IceCube array as an active veto to reject atmospheric muon background and high-energy ($>100$ GeV) neutrinos
\cite{SchulzEulerICRC}.
Standard neutrino oscillations are expected to affect only the low-energy sample. 
The high-energy reference sample provided high statistics outside the signal region and served to constrain systematic uncertainties.
The low-energy sample contained 719 events, while the high energy sample contained $39,638$ events after final cuts.


The directions of the neutrino-induced muon tracks in the high-energy sample were determined with the  standard  maximum 
likelihood muon track reconstruction of IceCube \cite{Ahrens:2003fg}.
For low-energy events, the same method was applied as an initial step. 
However, the standard hypothesis of a through-going track is not
appropriate at low energies. In a subsequent step, the length and end
points of the track are reconstructed and the likelihood of wether the
track started and/or stopped inside the detector volume is calculated \cite{SchulzEulerICRC}. 
Misreconstructed downward-going tracks originating from cosmic ray
muons are rejected by quality cuts on reconstruction variables like the
number of unscattered photons and the track likelihood. The resultant 
neutrino energy distributions of the two samples are shown in Fig.~\ref{fig:energy}.

The dominant background in the low-energy sample was misidentified (as track-like)  $\nu_e$ events, with a contribution of $10-15 \%$ 
as estimated from simulations.
The event selection has a non-zero efficiency for $\nu_\tau$ events, and some of the  $\nu_\mu$ that oscillate into $\nu_\tau$ 
will thus be retained in the sample.
We therefore included the $\nu_e$ background and the effect of $\nu_\tau$ appearance due to $\nu_\mu \rightarrow \nu_\tau$ in the analysis.
In $11$ days of simulated cosmic ray air shower data no events were found to pass the 
final cuts of the low-energy sample.  The dominant background in the high-energy sample was mis-reconstructed cosmic ray-induced muons contributing 
$5\%$. 

The resolution of the reconstructed zenith angle is an essential parameter given that the neutrino propagation length is proportional to the cosine of the zenith angle. 
The variation in zenith angles alters $L/E$ and thus the survival probability. 
The angular resolution of the low-energy sample was $8^\circ$ with respect to the neutrino direction, roughly independent of direction
and only slightly degrading with decreasing energy. 
The angle between the neutrino and the muon produced in a charged current interaction
amounts to about half of the measured zenith resolution.


We tested for an oscillation signal by evaluating the combined $\chi^2$ for histograms of the cosine of the reconstructed zenith angle for both
the high-energy and the low-energy sample. A bin size of 0.1 resulted in twenty bins.
Systematic uncertainties, considered via the covariance matrix $\sigma_{ij}$, give $\chi^2= \sum_{ij} R_i R_j \sigma^{-2}_{ij}$. 
Here, $R_i$ is the difference between the expected and measured rate in bin number $i$.
The covariance matrix is defined as $\sigma_{ij}^2=\delta_{ij}u_i u_j +\sum_k c_i^k c_j^k$ and depends
on uncorrelated (statistical) errors ($u_i$) in each bin as well as
on correlated (systematic) errors ($c_i^k=n_i^{\mathrm {std}}-n_i^{\mathrm{syst}, k}$). This approach implies the linear additive superposition of systematic errors. 
The term $n_i^{\mathrm{syst}, k}$ is the expected event rate in bin $i$ after modification of  the $k^{\mathrm{th}}$ systematic source of 
error by $1 \sigma$, and $n_i^{\mathrm{std}}$ is the default expectation in the same bin \cite{Fogli:2002pt}. Hence, the off-diagonal elements 
of the covariance matrix  reflect the bin-to-bin correlations of the systematic uncertainties, as expected. 
A set of sources of systematic uncertainties were considered explicitly and propagated by Monte Carlo simulation to the final selection level. Included are
the absolute sensitivity of the IceCube sensors ($\pm10\%)$ and the efficiency of the more sensitive DeepCore DOMs relative to the standard 
IceCube DOMs ($1.35\pm0.03$),
the optical parameters (scattering, absorption) of the ice as a detector medium where the uncertainty is estimated by 
the difference of the optical parameters obtained by the extraction methods \cite{Millenium} and \cite{Aartsen:2013rt}.
An additional systematic uncertainty for this analysis is associated with the atmospheric neutrino flux expectation given by~\cite{Honda:2006qj}.
Recent measurements of the spectrum of charged cosmic rays in the energy range 200 GeV to 100 TeV (e.g.~\cite{Yoon:2011aa})
indicate a flatter cosmic ray spectrum than that assumed in \cite{Honda:2006qj}. 
To reflect these new measurements we adjusted the neutrino spectrum by hardening the spectral index by 0.05.
Around this expectation we considered uncertainties in the absolute normalization ($\pm25\%$), the spectral index ($\pm0.05$) as well as the difference between
the calculations by \cite{Honda:2006qj} and \cite{Barr:2004br} for $\nu_\mu$ and for $\nu_e$.

\begin{figure}[hb!]
\begin{center}
\includegraphics[width=0.45\textwidth]{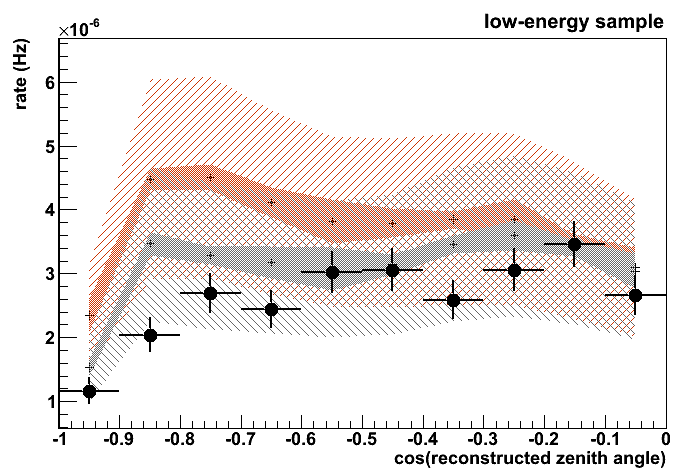}
\includegraphics[width=0.45\textwidth]{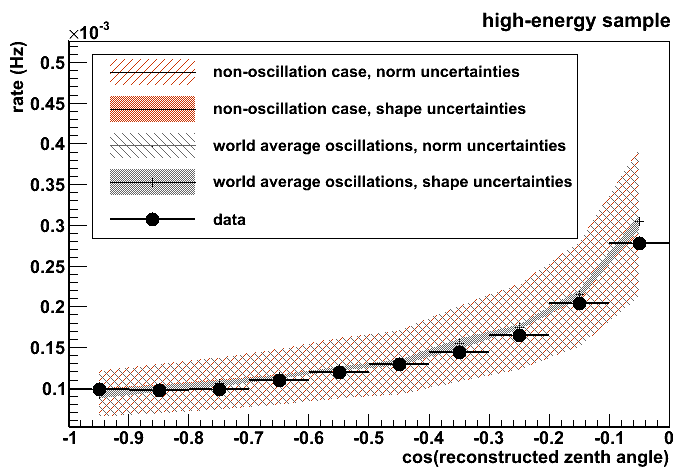}
\end{center}
\caption{\label{fig:nofit}Data and Monte Carlo expectation at world average oscillation parameters ($\sin^2(\theta_{23})=0.995$ and $\Delta m^2_{23}= 2.39 \cdot 10^{-3}$eV$^2$) 
\cite{Fogli:2008ig} and at the non-oscillation scenario 
for the low-energy sample and for the high-energy sample. For illustration purpose, systematic uncertainties are split into a fully correlated ("norm") part
and an uncorrelated ("shape") part. Both components are indicated by shaded error bands.
} 
\end{figure}

The $\chi^2$ was evaluated for two different physics hypotheses: a standard oscillation scenario
with the world average best fit parameters \cite{Fogli:2008ig}, and the non-oscillation scenario. The predicted zenith angle
distributions for both hypotheses are shown in Fig~\ref{fig:nofit} together with the data. We note good agreement between predictions and data
in both low- and high-energy (reference) samples. 
With $\Delta\chi^2=30$ between these hypotheses, 
a non-oscillation scenario is rejected with a 
p-value of $10^{-8}$ or $5.6\sigma$. The significance was evaluated with a toy Monte Carlo to account for deviations 
from a $\chi^2$ distribution since neither assumed hypothesis necessarily corresponds to the $\chi^2$ minimum.

\begin{figure}[hb!]
\begin{center}
\includegraphics[width=0.45\textwidth]{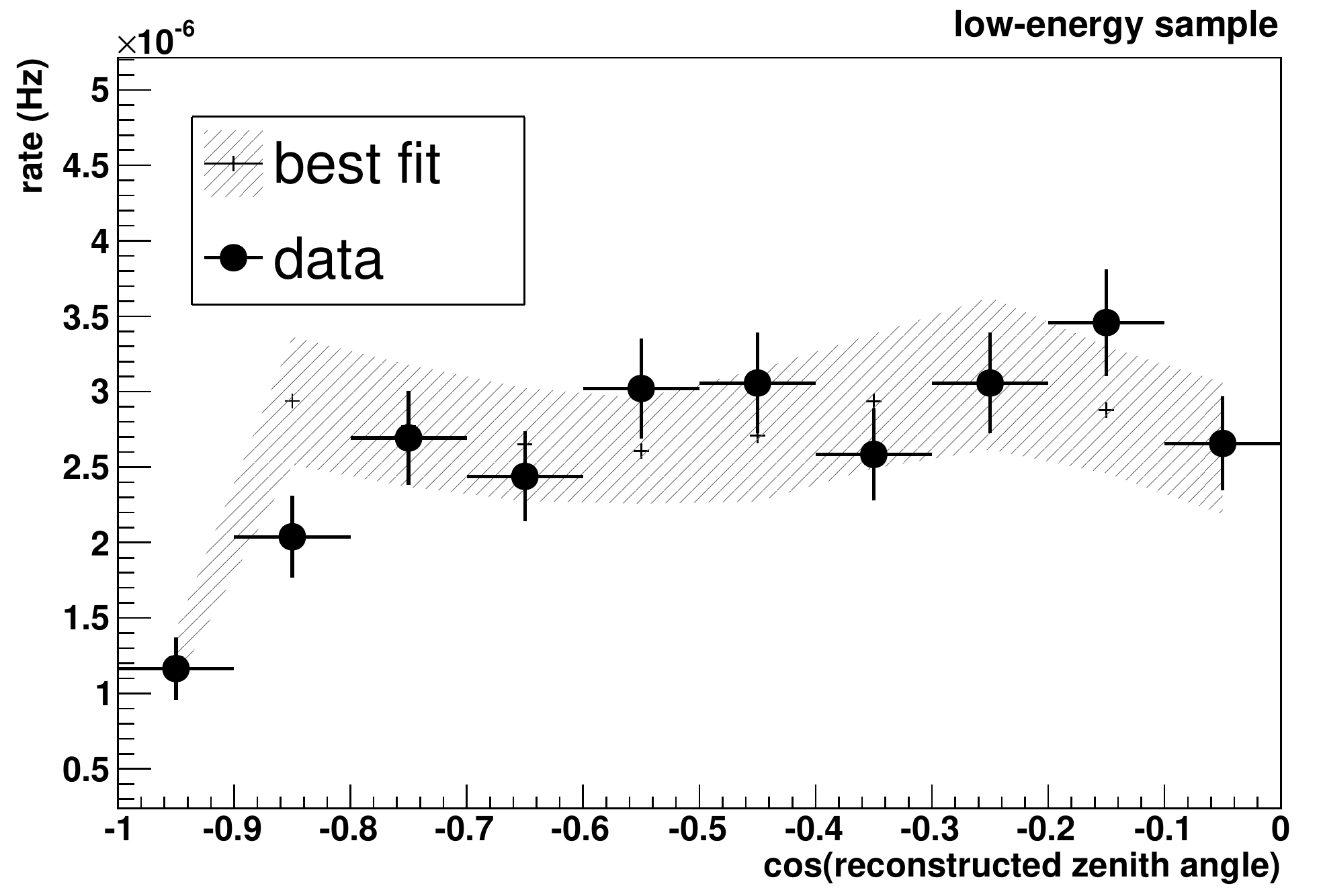}
\end{center}
\caption{\label{fig:bestfit}Data and Monte Carlo at best-fit oscillation parameters and pulls for the low-energy sample. 
The systematic uncertainty band is derived from the fit uncertainties of the pulls.} 
\end{figure}

\begin{table}
\begin{tabular}{l|c}
Systematic uncertainty&pull [std.\ deviations]\\
\hline
DOM efficiency & 0.32\\
Ice model & -0.12\\
Atm.\ flux model & -0.59\\
Normalization & -0.82\\
CR index / cross section & 0.42 \\
Relative efficiency of DeepCore DOMs & -0.01\\
Normalization of $\nu_e$& -0.53
\end{tabular}
\caption{\label{table:pulls}Pulls on the systematic uncertainties at best fit value of $\Delta m^2_{23}=2.3\cdot10^{-3} \mathrm{eV}^2$ and $\sin^2(2 \theta_{23})=1$.}
\end{table}
\label{results}
\begin{figure}[htb!]
\begin{center}
\includegraphics[width=0.4\textwidth]{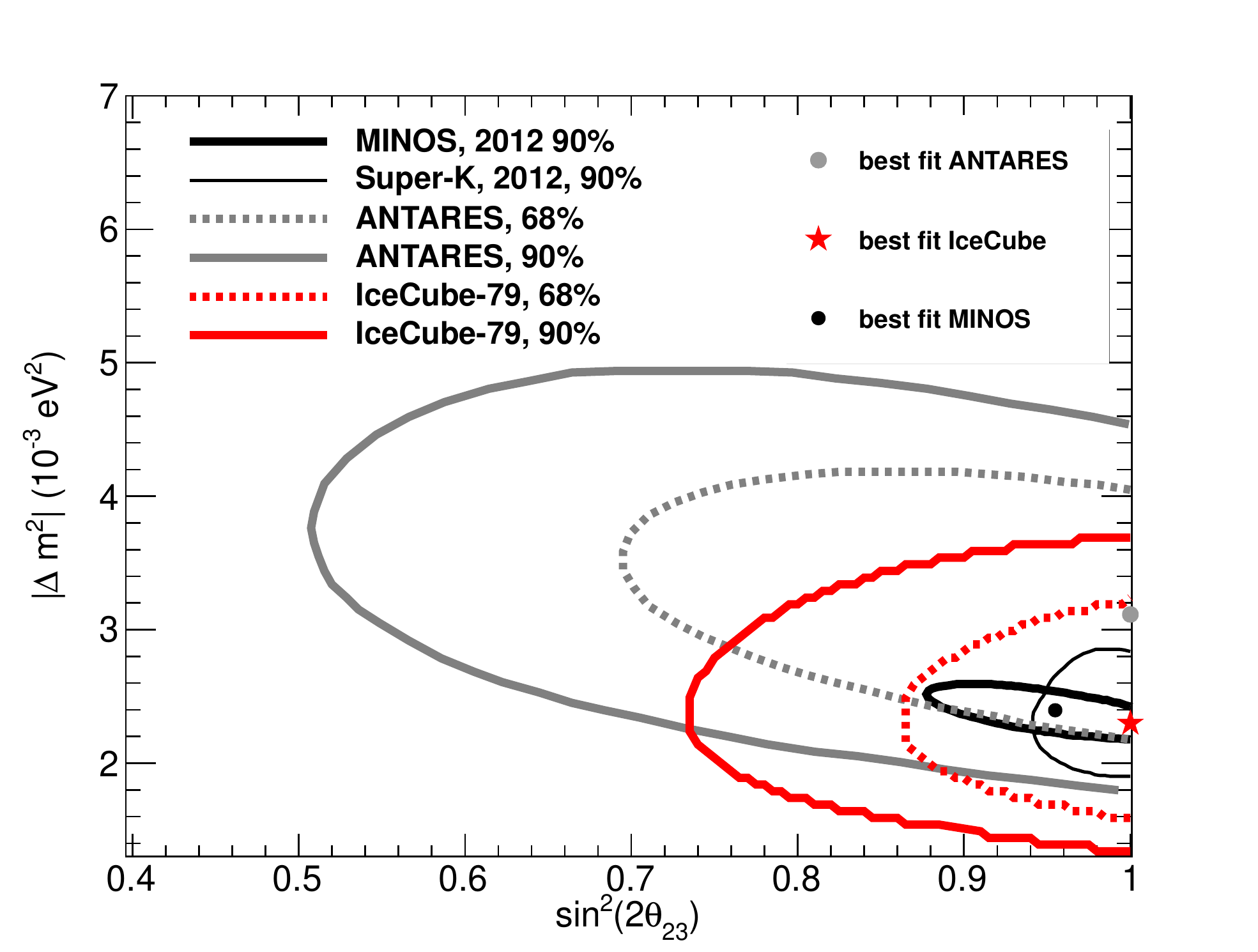}
\end{center}
\caption{\label{fig:contourlines} Significance contours for the presented atmospheric neutrino oscillation analysis, compared with the 
results of ANTARES \cite{AdrianMartinez:2012ph}, MINOS \cite{Minos_Neutrino} and SuperKamiokande \cite{Itow_Neutrino}.}
\end{figure}

The $\chi^2$ was also evaluated as a function of the oscillation parameters, using the \emph{pull} method outlined in \cite{Fogli:2002pt}. The
parameters considered as sources of systematic uncertainty in the Monte Carlo
prediction were fitted simultaneously with the oscillation parameters. The expected zenith angle distribution at best fit (oscillation parameters and systematic uncertainties) 
are shown in Fig.~\ref{fig:bestfit} for the low-energy sample.
The best-fit systematics parameters (represented by the \emph{pulls}) are listed in Table~\ref{table:pulls}. All pulls were within 
the  $1\sigma$ band, indicating a self-consistency of the analysis.
The best fit oscillation parameters are given by $\Delta m^2_{23}=2.3\cdot10^{-3}$eV$^2$ 
and $\sin^2(2 \theta_{23})=1$, with  $\chi^2=15.7$ and 18 degrees of freedom (20 bins, 2 fitted parameters). 


The two-dimensional confidence regions of the oscillation parameters were determined from the $\Delta\chi^2$ around the best fit with
two degrees of freedom. 
The resultant regions are shown in Fig.~\ref{fig:contourlines} together with results from other experiments 
\cite{Minos_Neutrino, Itow_Neutrino}.
A full Monte Carlo ensemble test, sampling true values for the considered sources of systematic errors according to Gaussian statistics and Poisson fluctuations in the 
observed bin counts, was used to map the test statistics.
A slight overcoverage at $78\%$ was found for the $1 \sigma$ contour, related to the proximity of the mixing angle to 
the maximum mixing boundary, i.e.~the obtained contours are conservative.
The confidence regions for the individual parameters were determined by marginalization analogous to a profile likelihood method. We obtain 
$68\%$ confidence intervals of $\Delta m^2_{23}= (2.3^{+0.5}_{-0.6})\cdot 10^{-3}$ eV$^2$ and $\sin^2(2 \theta_{23})>0.93$ using a $\Delta\chi^2$ with one degree of freedom.

This analysis of IceCube data has provided the first significant detection ($>5\sigma$) of atmospheric neutrino oscillations at energies near the 
$25$ GeV oscillation maximum for vertical events. The measured oscillation parameters are in good agreement
with results from other experiments that have measured the atmospheric oscillation parameters with high resolution at lower energies. 
Hence, these measurements agree with the theoretical predictions of the standard three-neutrino 
flavor oscillation framework. 
Significant future improvements in our sensitivity to atmospheric neutrino oscillations are expected by 
the application of new reconstruction methods
that are more efficient at the lowest energies covered by DeepCore.
We expect that the rate of detected atmospheric neutrinos 
near the $25$ GeV oscillation maximum will be increased significantly. These
higher statistics will lead to tighter constraints on the oscillation parameters with IceCube.
Furthermore, the inclusion of the reconstructed energy as a second analysis variable will improve the constraints in particular on $\Delta m^{2}_{23}$.
Additionally, improvement is expected from the inclusion of the two final DeepCore strings which 
started taking data in May 2011.

We acknowledge the support from the following agencies:
U.S. National Science Foundation-Office of Polar Programs,
U.S. National Science Foundation-Physics Division,
University of Wisconsin Alumni Research Foundation,
the Grid Laboratory Of Wisconsin (GLOW) grid infrastructure at the University of Wisconsin - Madison, the Open Science Grid (OSG) grid infrastructure;
U.S. Department of Energy, and National Energy Research Scientific Computing Center,
the Louisiana Optical Network Initiative (LONI) grid computing resources;
Natural Sciences and Engineering Research Council of Canada,
WestGrid and Compute/Calcul Canada;
Swedish Research Council,
Swedish Polar Research Secretariat,
Swedish National Infrastructure for Computing (SNIC),
and Knut and Alice Wallenberg Foundation, Sweden;
German Ministry for Education and Research (BMBF),
Deutsche Forschungsgemeinschaft (DFG),
Helmholtz Alliance for Astroparticle Physics (HAP),
Research Department of Plasmas with Complex Interactions (Bochum), Germany;
Fund for Scientific Research (FNRS-FWO),
FWO Odysseus programme,
Flanders Institute to encourage scientific and technological research in industry (IWT),
Belgian Federal Science Policy Office (Belspo);
University of Oxford, United Kingdom;
Marsden Fund, New Zealand;
Australian Research Council;
Japan Society for Promotion of Science (JSPS);
the Swiss National Science Foundation (SNSF), Switzerland.
\bibliography{IC79_oscillations}

\end{document}